\documentclass[12pt,thmsa]{article}
\usepackage{amssymb}

\begin{document}

\title{Parametrization of dark energy equation of state }
\author{Vinod B. Johri$^{1,2}$ and P.K.Rath$^{3}$ \\
$^{1}${\small Department of Mathematics and Astronomy, University of
Lucknow, Lucknow-226007, India. }\\
{\small E-mail:vinodjohri@hotmail.com }\\
$^{2}${\small William I. Fine Theoretical Physics Institute, University of
Minnesota, Minneapolis, MN 55455, USA}\\
$^{3}${\small Department of Physics, University of Lucknow, Lucknow-226007,
India}\\
{\small E-mail: pkrath\_lu@yahoo.co.in}}
\date{}
\maketitle

\begin{abstract}
A comparative study of various parametrizations of the dark energy equation
of state is made. Astrophysical constraints from LSS, CMB and BBN are laid
down to test the physical viability and cosmological compatibility of these
parametrizations. A critical evaluation of the 4-index parametrizations
reveals that Hannestad-M\"{o}rtsell as well as Lee parametrizations are
simple and transparent in probing the evolution of the dark energy during
the expansion history of the universe and they satisfy the LSS, CMB and BBN
constraints on the dark energy density parameter for the best fit values.
\end{abstract}

\section{Introduction\label{sect-intro}}

The recently discovered 16 Type Ia supernovae (SN Ia) with the Hubble
Telescope by Riess et al. \cite{ries04} provide a distinct scenario and
conclusive evidence of the decelerating universe in the past $(z>1)$
evolving into the present day accelerating universe. Thus, the existence of
dark energy, which accelerates the cosmic expansion, has been firmly
established and the present magnitude of its energy density has been
precisely determined \cite{wang04a}. The focus is now on the evolution of
dark energy and its equation of state with the cosmic expansion. The
simplest and the most natural candidate for dark energy is the cosmological
constant $\Lambda $ with a constant energy density $\rho _{\Lambda }$ and a
fixed equation of state parameter $w=-1$. But the recent analysis of the SN
Ia data \cite{alam04,hute04} indicates that the time varying dark energy
gives a better fit to observational data than a cosmological constant which
has only one free parameter $\Omega _{M}^{0}$. Assuming the time varying
equation of state, an important issue arises whether the dynamical dark
energy can cross the phantom divide line (PDL) $w=-1$ during evolution? To
put it more explicitly, can quintessence transform into phantom dark energy
during the course of its evolution? There are divergent views on this
question [5-8] but it is found that such parametrizations which cross the
PDL seem to provide slightly better fit to observational data \cite{stef05}.

As regards the rapidity of the evolution of dark energy, there are two
contrasting views at present. Riess et al. \cite{ries04} and Jassal et al. 
\cite{jass04} have argued that the current SN Ia observational data is
inconsistent with the rapid evolution of $w(z)$. On the other hand, Bassett
et al. \cite{bass02} and Corasaniti et al. \cite{cora04} contend that the
inconsistency arises on account of the inadequacy of 2-index
parametrization.. They claim that the rapid variation in $w(z)$ in fact
provides a better fit to the SN Ia `gold set' observations, even after
including CMB and large scale structure data \cite{cora04}. According to our
analysis, the rapidity of variation $|\frac{dw}{dz}|$, apart from other
factors, depends on the absolute value of $|w|$. For quintessence models $%
-1<w<-\frac{1}{3}$, whereas $w$ has no lower bound for phantom models.
Hence, the equation of state of dark energy varies more rapidly for phantom
models with large $|w|$ than for quintessence models, other factors
remaining unchanged. The fact that the observational data by Caldwell et al%
\textit{.} \cite{cald03} including SN Ia and galaxy clustering, shows a bias
towards phantom models might explain why rapid variation in $w(z)$ provides
a better fit to observational data.

It is difficult to know the exact functional form of the equation of state
parameter $w(z)$ observationally as such different parametric forms namely
one index parametrization by Gong et al. \cite{gong05}, two index
parametrization [9, 14-21], four index parametrization by Hannestad et al. 
\cite{hann04} and Lee \cite{lee05}, power law parametrization \cite
{well02,well01,asti01}, more complicated forms [26-29] and non-parametric
forms Bassett et al. \cite{bass02}, Corasaniti et al. \cite{cora04,cora02}
and Sahni et al. \cite{sahn03} of $w(z)$ have been used to simulate the
behavior of evolving equation of state of dark energy and comparison made
with the SN Ia observations. Alternatively, the dark energy density also can
be parametrized as power law expansion [31-36] or piece wise constant
parametrization \cite{wang04a,hute04,wang04b,card04}. The SN Ia observations
essentially measure the luminosity distance $d_{L}(z)$ which when compared
with the theoretical parametric values, yields the best fit values of the
parameters. But it may be stressed that these observational tests alone are
not adequate to distinguish between the cosmological constant and the
dynamical dark energy. If we start with the assumption $w=constant$, the
observational data favors $w=-1$. In case of dynamical dark energy, the
current observational data does not yield any conclusive constraint over the
equation of state parameter $w(z)$. In fact, large downward deviations of
the equation of state parameter from $w=-1$ favor phantom dark energy.
Therefore, we need to devise some new tests for probing dark energy.
Odintsov and Nojiri \cite{odin04} have discussed the thermodynamics of the
evolution of dark energy and have stressed that the phantom stage might be a
transient period in cosmic evolution. So far we have been exploring the
properties of dark energy (its equation of state and the density parameter $%
\Omega _{X}$) using only cosmological observations. A new approach might be
to use astrophysical constraints to probe the various dark energy
parametrizations of $w(z)$ as discussed in this paper.

For making a comparative study of various parametrizations and their
viability, we need some criterion. In our previous papers [40-42], we
introduced the concept of `integrated tracking' and outlined certain
astrophysical constraints to be satisfied by the dark energy fields for
realistic tracking. Recently, Upadhye et al. \cite{upad04} have discussed
the dark energy constraint from CMB data at decoupling epoch $z=1100$. The
astrophysical constraints discussed below arise from well established
cosmological observations at different redshifts.

(A) LSS Constraint

During the galaxy formation era ($1\ <\ z\ <\ 3$), dark energy density must
be sub-dominant to matter density; accordingly $\Omega _{X}\;<\;0.5$ \cite
{free03}.

(B) CMB Constraint

Caldwell \cite{cald04} and Upadhye et al\textit{.} \cite{upad04} have argued
that $(\Omega _{X})_{dec}\;<\;0.1$ at $z\;=\;1100$. It imposes a high
redshift upper bound $\Omega _{X}$ for any viable parametrization of the
equation of state parameter $w(z)$.

(C) BBN Constraint

The presence of dark energy until nucleosynthesis epoch, should not disturb
the observed Helium abundance in the universe which is regarded as the
foundation stone of the Big Bang Theory. According to Johri \cite{johr02p} $%
(\Omega _{X})_{BBN}\;<\;0.14$ at $z\;=\;10^{10}$ whereas the latest analysis
of Cyburt \cite{cybu04} constrains $(\Omega _{X})_{BBN}\;<\;0.21$ at $%
z\;=\;10^{10}$. This puts a stringent high redshift limit on dark energy.

In this paper, we have discussed the cosmological compatibility and the
range of validity of six dark energy parametrizations, the first four
involving 2-parameters and the fifth and sixth involving 4-parameters. More
parameters mean more degrees of freedom for adaptability to observations, at
the same time more degeneracies in the determination of parameters. All the
parametrizations are critically examined on the basis of the above
constraints to test how faithfully they represent the behavior of dynamical
dark energy in the expanding universe.

The outline of the paper is as follows. In section~\ref{sect-equations}, we
have discussed the expansion history of the universe. Assuming a spatially
flat universe, the field equations give the Hubble expansion and
deceleration parameter in terms of function $f(z)$ which involves integral
of the varying equation of state parameter of dark energy. In section 3, we
have discussed 2-index parametrization models namely linear red shift
parametrization \cite{hute00,well02}, Chevallier-Polarski-Linder
parametrization \cite{pola01,lind03}, Jassal-Bagla-Padmanabhan
Parametrization \cite{jass04} and Upadhye-Ishak-Steinhardt parametrization 
\cite{upad04}. In section 4, we have critically examined
Hannestad-M\"{o}rtsell parametrization \cite{hann04} and Lee's model \cite
{lee05} which involve 4 parameters imparting four degrees of freedom to
choose them. In section 5, we use interpolation technique to study the
behavior of the dark energy under the assumption of slowly varying equation
of state. In section 6, we conclude with some remarks on parametrization
methods.

\section{Expansion history of the universe\label{sect-equations}}

Assuming a spatially flat ($k=0$) Friedmann universe, the field equations
are 
\begin{equation}
H^{2}\;\;=\;\;H_{o}^{2}[\Omega _{M}^{o}(1+z)^{3}+\Omega _{X}^{o}f(z)]
\end{equation}
and 
\begin{equation}
\frac{2q-1}{3}\;\;=\;\;w(z)\Omega _{X}
\end{equation}
where $H=\frac{\dot{a}}{a}$ is the Hubble constant, $q=-\frac{\ddot{a}}{%
aH^{2}}$ is the deceleration parameter, $w(z)=\frac{p_{X}}{\rho _{X}}$ is
the equation of state of dark energy and $\Omega _{X}=1-\Omega _{M}$ is the
cosmic dark energy density parameter. The dark energy density parameter is
defined as 
\begin{equation}
\Omega _{X}=\frac{\rho _{X}}{\rho _{C}}
\end{equation}
with 
\begin{equation}
\rho _{X}=\rho _{X}^{o}f(z)
\end{equation}
and 
\begin{equation}
f(z)\;\;=\;\;\exp \left[ 3\int_{0}^{z}{\frac{1+w(z^{\prime })}{1+z^{\prime }}%
dz^{\prime }}\right]
\end{equation}
We can test any parametrization by taking the best fit values to the
observed luminosity distance of SN Ia given by 
\begin{equation}
d_{L}(z)\;\;=\;\;(1+z)\int_{0}^{z}{\frac{dz^{\prime }}{H(z^{\prime })}}
\end{equation}
and applying constraints (A), (B) and (C) to check its compatibility with
the cosmological observations. In Table 1, we have tabulated the dark energy
parameters $\Omega _{x}$\ at LSS formation, decoupling and BBN corresponding
to $z=1-3$, $1100$\ and $10^{10}$\ respectively. The transition redshift may
also be derived from the best fit values.

\section{Two\ index\ parametrization}

\textbf{I. Linear-redshift parametrization}

\vspace{2mm} The equation of state parameter $w(z)$ is given by Huterer et
al. \cite{hute00} and Weller et al. \cite{well02} as 
\begin{equation}
w(z)\;\;=\;\;w_{o}+w^{\prime }z,\quad \quad w^{\prime }=\Big( \frac{dw}{dz}%
\Big)_{z=0}
\end{equation}
Inserting Eq. (7) for $w(z)$ in Eq.(3), we get 
\begin{equation}
\Omega _{X}(z)=\left[ 1+\frac{\Omega _{M}^{0}}{\Omega _{X}^{0}}%
(1+z)^{-3(w_{0}-w^{\prime })}\exp (-3w^{\prime }z)\right] ^{-1}
\end{equation}
It has been used by Riess et al. \cite{ries04} for probing SN Ia
observations at $z<1$. The best fit values to SN Ia `gold set' data \cite
{dicu04} are $w_{o}=-1.4$, $w^{\prime }=1.67$ and $\Omega _{M}^{o}=0.30$.
Hence, the parameters favor dark energy of phantom origin. In the distant
past ( $z>>1$), $w\rightarrow \infty $ and $\Omega _{X}\rightarrow 1$ and in
the distant future ($1+z\rightarrow 0$), $w=w_{0}-w^{\prime }{}=-3.07$ and $%
\Omega _{X}\rightarrow 1$. At the transition red shift $z_{T}=0.39$, we get $%
(\Omega _{X})_{T}=0.443$. The parametrization is not viable as it diverges
for $z>>1$ and therefore incompatible with the constraints (B) and (C).

\vspace{2mm} \textbf{II. Chevallier-Polarski-Linder parametrization}

\vspace{2mm} The equation of state parameter $w(z)$ is given as \cite
{pola01,lind03} 
\begin{equation}
w(z)\;\;=\;\;w_{o}+w_{1}\frac{z}{1+z}
\end{equation}
On differentiation, $\frac{dw}{dz}\;=\;\frac{w_{1}}{(1+z)^{2}}$. It
indicates rapid variation of $w(z)$ at $z=0$ which goes on decreasing with
increasing $z$. Eq.(3) gives 
\begin{equation}
\Omega _{X}(z)=\left[ 1+\frac{\Omega _{M}^{0}}{\Omega _{X}^{0}}%
(1+z)^{-3(w_{0}+w_{1})}\exp (\frac{3w_{1}z}{1+z})\right] ^{-1}
\end{equation}
The best fit values of $w_{o},$ $w_{1}$ and $\Omega _{m}^{0}$ to SN Ia `gold
set' data \cite{gong04a,dicu04} turn out to be $-1.6,$ $3.3$ and $0.3$
respectively$.$ At $z=0$, one has $w(z)=w_{0}=-1.6$.$\;$and the parameters
suggest that the dark energy is of phantom origin. For $z>>1$, $w\rightarrow
w_{0}+w_{1}=1.7$ and $\Omega _{X}\rightarrow 1$. In the distant future ($%
1+z\rightarrow 0$), $w\rightarrow -\infty $ and $\Omega _{X}\rightarrow 1$.
The dark energy parameter $(\Omega _{X})_{T}=0.452$ at the transition red
shift $z_{T}=0.35$. The CMB\ and BBN constraints for $\Omega _{X}$ are
violated. Hence, it is incompatible with observation.

\vspace{2mm} \textbf{III. Jassal-Bagla-Padmanabhan parametrization}

\vspace{2mm} The equation of state parameter $w(z)$ is given by Jassal et
al. \cite{jass04} as

\begin{equation}
w(z)\;\;=\;\;w_{o}+\frac{w_{1}z}{(1+z)^{2}}
\end{equation}
As $\frac{dw}{dz}=\frac{w_{1}(1-z)}{(1+z)^{3}}$, $w(z)$ increases from $z=0$
to $z=1$ thereafter starts decreasing. Eq.(3) gives 
\begin{equation}
\Omega _{X}(z)=\left[ 1+\frac{\Omega _{M}^{0}}{\Omega _{X}^{0}}%
(1+z)^{-3w_{0}}\exp \left\{ -\frac{3w_{1}z^{2}}{2(1+z)^{2}}\right\} \right]
^{-1}
\end{equation}
The best fit values to SN Ia `gold set' data are $w_{o}=-1.9,w_{1}=6.6$ and $%
\Omega _{M}^{o}=0.3$. One has $w(z)=w_{0}=-1.9$ at $z=0$. Thus, the
parameters suggest that the dark energy is of phantom origin. In the distant
past for $z>>1$, $w(z)\rightarrow w_{0}=-1.9$ and $\Omega _{X}\rightarrow 0$%
. In the distant future ($1+z\rightarrow 0$), $w\rightarrow -\infty $ and $%
\Omega _{X}\rightarrow 1$. At $z_{T}=0.3$, one gets $(\Omega _{X})_{T}=0.467$%
. Further, it satisfies all the three observational constraints (A), (B) and
(C).

\vspace{2mm} \textbf{IV. Upadhye-Ishak-Steinhardt parametrization}

\vspace{2mm} Upadhye et al. have parametrized $w(z)$ as \cite{upad04} 
\begin{equation}
w(z)=\left\{ 
\begin{array}{lll}
w_{0}+w_{1}z & if & z<1 \\ 
w_{0}+w_{1} & if & z\geqslant 1
\end{array}
\right.
\end{equation}
The dark energy density contrast is given by 
\begin{equation}
\Omega _{X}(z)=\left\{ 
\begin{array}{c}
\begin{array}{lllll}
\left[ 1+\frac{\Omega _{M}^{0}}{\Omega _{X}^{0}}(1+z)^{-3(w_{0}-w_{1})}\exp
(-3w_{1}z)\right] ^{-1} &  &  & if & z<1
\end{array}
\\ 
\begin{array}{lllll}
\left[ 1+\frac{\Omega _{M}^{0}}{\Omega _{X}^{0}}%
(1+z)^{-3(w_{0}+w_{1})}e^{-3w_{1}\left( 1-2\ln 2\right) }\right] ^{-1} &  & 
if &  & z\geqslant 1
\end{array}
\end{array}
\right.
\end{equation}
The best fit values of parameters for SN Ia `gold set', galaxy power
spectrum and CMB power spectrum data are $w_{0}=-1.38$, $w_{1}=1.2$ and $%
\Omega _{M}^{0}=0.31$. The transition red-shift $z_{T}$ turns out to be $%
0.44 $. At $z=0$, one has $w=w_{0}=-1.38$ and this suggests that the dark
energy has phantom origin. In the distant past ($z>>1$), $w\rightarrow -0.18$
and $\Omega _{X}\rightarrow 0$ and in the distant future ($1+z\rightarrow 0$%
), $w=w_{0}-w_{1}=-2.58$ and $\Omega _{X}\rightarrow 1$. One gets $(\Omega
_{X})_{T}=0.392$ at transition redshift. The parameterization also satisfies
the LSS, CMB and BBN\ constraints.

\section{Four index parametrization}

\textbf{I. Hannestad-M\"{o}rtsell parametrization}

\vspace{2mm} Let us now consider Hannestad parametrization \cite{hann04}
which involves 4-parameters. 
\begin{eqnarray}
w(z)\;\; &=&\;\;w_{o}w_{1}\frac{a^{p}+a_{s}^{p}}{w_{1}a^{p}+w_{o}a_{s}^{p}}\;
\nonumber \\
\; &=&\;\;\frac{1+\big(\frac{1+z}{1+z_{s}}\big)^{p}}{w_{o}^{-1}+w_{1}^{-1}%
\big(\frac{1+z}{1+z_{s}}\big)^{p}}
\end{eqnarray}
where $w_{0}$ and $w_{1}$ are the asymptotic values of $w(z)$ in the distant
future ($1+z\rightarrow 0$) and in the distant past ($z\rightarrow \infty $)
respectively. The $a_{s}$ and $p$ are the scale factor at the change over
and the duration of the change over in $w$ respectively.

It follows that the equation of state at the present epoch $(z=0)$ is given
by 
\begin{equation}
w^{*}\;\;=\;\;\frac{w_{o}w_{1}(1+a_{s}^{p})}{w_{1}+w_{o}a_{s}^{p}}
\end{equation}
Differentiating Eq.(15) with respect to $z$, we get 
\begin{equation}
w^{-1}\frac{dw}{dz}\;=\;\frac{(w_{1}-w_{o})p(1+z)^{p-1}a_{s}^{p}}{%
[1+(a_{s}/a)^{p}][w_{1}+w_{o}(a_{s}/a)^{p}]}
\end{equation}
Hence, the equation of state parameter of dark energy varies with the
redshift unless $w_{1}=w_{o}$ or $w(z)$ takes asymptotic values $%
1+z\rightarrow \infty $ or $1+z\rightarrow 0$. The variation gradient $|%
\frac{dw}{dz}|$ varies directly as $|w|,$ $|w_{1}-w_{o}|,$ $p$ and inversely
as $a_{s}^{p}(1+z)^{p}$.

Inserting for $w(z)$ from Eq.(15) into Eq.(3), we get 
\begin{equation}
\Omega _{X}(z)=\left[ 1+\frac{\Omega _{M}^{0}}{\Omega _{X}^{0}}%
(1+z)^{-3w_{1}}\left\{ \frac{w_{1}+w_{0}a_{s}^{p}\left( 1+z\right) ^{p}}{%
\left( w_{1}+w_{0}a_{s}^{p}\right) \left( 1+z\right) ^{p}}\right\} ^{\frac{%
-3\left( w_{0-}w_{1}\right) }{p}}\right] ^{-1}
\end{equation}
Taking the best fit values for the combined CMB, LSS and SN Ia data \cite
{hann04}, $w_{o}=\;-1.8,$ $w_{1}=\;-0.4$ $,$ $p\;=\;3.41$ and $a_{s}=0.50$
with a prior $\Omega _{M}^{o}=0.38$, one has $w(z=0)=-1.38$ at $z=0.$ So,
the parameters suggest that the dark energy is of phantom origin. In the
distant past ( $z>>1$), $w(z)\rightarrow w_{1}=-0.4$ and $\Omega
_{X}\rightarrow 0$. In the distant future ($1+z\rightarrow 0),$ $%
w\rightarrow w_{0}=-1.8$ and $\Omega _{X}\rightarrow 1$. At transition
redshift $z_{T}=0.39$, the dark energy parameter $(\Omega _{X})_{T}=0.333$.
Further, all the observational constraints are satisfied by this
parametrization.

\vspace{2mm} \textbf{II. Lee parametrization}

\vspace{2mm} The equation of state parameter $w(z)$ is parametrized by Lee
as \cite{lee05} 
\begin{equation}
w(z)=w_{r}\frac{w_{0}\exp (px)+\exp (px_{c})}{\exp (px)+\exp (px_{c})}
\end{equation}
where 
\begin{equation}
x=\ln a=-\ln (1+z)
\end{equation}
and the symbols $w_{r}$ and $x_{c}$ have the usual meaning as in \cite{lee05}%
. The dark energy parameter $\Omega _{X}(z)$ is given by 
\[
\Omega _{X}(z)=\left[ 1+\frac{\Omega _{M}^{0}}{\Omega _{X}^{0}}%
(a+a_{eq})\left( \frac{a^{p}+a_{c}^{p}}{1+a_{c}^{p}}\right) ^{-4/p}\right]
^{-1}
\]
The parameters $w_{r}$ is chosen to be $1/3$ using the tracking condition
and $w_{0}$ is taken as $-3$. The other parameters are obtained by analyzing
the separation of CMB\ peaks and the time variation of the fine structure
constant. For $\Omega _{M}^{0}=0.27$ and $x_{c}=-2.64$, the range of $p$ is
taken as $1.5\leq p\leq 3.9$. At $z=0$, one has $-0.975\leq w(z=0)\leq -1.0$
corresponding to $1.5\leq p\leq 3.9$. Hence, the parameters suggest that the
dark energy is of quintessence origin. For $p=1.5$ and $3.9$, the dark
energy parameter $\Omega _{X}$ is 0.353 and 0.333 at the transition red
shift $z_{T}=0.74-0.76$ respectively. In the distant past for $z>>1$, $%
w(z)\rightarrow w_{r}=0.333$ and $\Omega _{x}\rightarrow \left[ 1+\frac{%
\Omega _{M}^{0}}{\Omega _{X}^{0}}\frac{a_{eq}}{a_{c}^{p}}\right] ^{-1}$. In
the distant future, $w\rightarrow w_{r}w_{0}=-1$ and $\Omega _{x}\rightarrow
1$. In this case, all the three observational constraints (A), (B) and (C)
are satisfied.

\section{Interpolation of w(z) for slowly varying equation of state}

If we go by the analysis of Riess et al. \cite{ries04} and Jassal et al. 
\cite{jass04}, the current SN Ia data is inconsistent with rapid evolution
of dark energy. Therefore, we can use `integrated tracking' and
interpolation techniques applicable to slow time varying equation of state 
\cite{johr02p}, It was shown \cite{johr02c} that the scalar fields with
slowly varying equation of state which satisfy tracking criteria, are
compatible with astrophysical constraints (A) and (C) outlined under
`integrated tracking'. According to our analysis \cite{johr02p}, based on
integrated tracking, we have 
\[
\begin{array}{lll}
\Omega _{X}\;\leq \;\;0.14, & w\simeq -0.035\;\;\; & at\;\;z\;=\;10^{10} \\ 
\Omega _{X}\;\;=\;\;0.66, & w\;=\;-0.77\;\; & at\;\;z\;\;=\;0 \\ 
\Omega _{X}\;\;=\;\;0.5, & w\;\;=\;-0.66\;\; & at\;\;z_{T}\;=\;0.414
\end{array}
\]
The above data is consistent with the transition redshift given by Riess et
al. \cite{ries04} and $(\Omega _{X})_{BBN}$ given by Cyburt et al. \cite
{cybu04}.

\section{Conclusions}

We have investigated the behavior of dynamical dark energy by taking
parametric representations of the equation of state. Out of the 2-parameter
models of the equation of state, Jassal-Bagala-Padmanabhan parametrization
and Upadhye-Ishak-Steinhardt parametrizations are found to be valid and
compatible with the astrophysical constraints over a wide range of redshift.

Bassett et al. and Corasaniti et al. \cite{bass02,cora04} have discussed the
limitations of 2-index parametrizations. Most of them track well at low
redshifts $z\leq 0.2$ but if we explore dark energy for regions $z>1$ and $%
w<<-1$ (beyond quintessence models), very rapid variation in $w(z)$ can be
found. Higher order parametrizations are more suitable for probing the
nature of dark energy and its evolution since more parameters give more
freedom to fit in observational data but at the same time it gives rise to
more degeneracies in the determination of the parameters,

The Hannestad-M\"{o}rtsell model and Lee 4-parameter model of the equation
of state provide well- behaved representations of dark energy evolution over
a long range of redshift; they admit rapid variation of the equation of
state as well. Since, there is no lower bound on $w(z)$ for the phantom
models, $\Omega _{X}(z)$ also decreases steeply in the early universe if $%
1+w_{1}$ is negative.

\section*{Acknowledgments}

This work was done under the DST (SERC) Project, Govt. of India. One of the
authors VBJ gratefully acknowledges the hospitality extended by W. I. Fine
Theoretical Physics Institute, University of Minnesota, useful discussions
with Keith Olive and valuable help of Roman Zwicky in the compilation of
this manuscript.

\end{document}